# Real-time spectroscopic monitoring of continuous synthesis of zinc oxide nanostructures in femtosecond laser fabricated 3D microfluidic microchannels with integrated on-chip fiber probe array

Miao Wu,[a, b] Xin Li,[a*] Di-Feng Yin,[c] Wei Chen,[a, b] Jia Qi,[a, b] Ming Hu,[a] Jian Xu,[a, b] and Ya Cheng[a, b, c] **

a State Key Laboratory of Precision Spectroscopy, Engineering Research Center for Nanophotonics &Advanced Instrument, Ministry of Education, School of Physics and Electronic Science, East China Normal University, Shanghai, 200241, China

b XXL –The Extreme Optoelectromechanics Laboratory, School of Physics and Electronic Science, East China Normal University, Shanghai, 200241, China

c State Key Laboratory of High Field Laser Physics and CAS Center for Excellence in Ultra-intense Laser Science, Shanghai Institute of Optics and Fine Mechanics (SIOM), Chinese Academy of Sciences (CAS), Shanghai, 201800, China

Email：xli@phy.ecnu.edu.cn, ya.cheng@siom.ac.cn

# Abstract

Materials synthesis in a microfluidic environment enables the flexible and controllable production of various types of nanostructures which are of great potential in the fields of chemistry, environmental science, bioengineering, and medicine. Here, we demonstrate on-chip simultaneous continuous-flow synthesis and in-situ spectrum diagnosis of zinc oxide (ZnO) nanomaterials using a femtosecond-fabricated three-dimensional microchannel reactor integrated with an array of optical fiber probes. The microchannel reactor including 3D concentration gradient generators followed by 3D micromixing units provides high-efficiency manipulation of reactants with different concentrations as well as parallel reaction dynamics in an autonomous manner. The integrated optical fiber probe array allows precise and parallel spectropic detection in different microchannels with high spatial and temporal resolutions for screening the synthetic conditions. The synthesized ZnO nanostructures can be tailored in size, shape, and morphology by tuning the flow rates and reactant concentrations based on the spectroscopic signals detected with the fiber probe array.

**Introduction**

Continuous flow synthesis based on microchannel reactors has been recognized as an enabling technology for the rapid production of fine chemicals and advanced materials owing to their advantages such as high reaction efficiency, intrinsic safety, and feasibility in scaling-up production with the ability to perform parallel reactions.[1-5] In the continuous flow synthesis, the configurations, and functionalities of microchannel structures are of critical importance to manipulate and monitor the microfluidic behaviors for high-performance mixing of reactants,[6] heat and mass transfer,[7,8] and separation of products.[9-14] Currently, there is an increasing demand for developing microchannel reactors with three-dimensional (3D) configurations since they provide superior capabilities for fluid manipulation compared with conventional 2D conditions in confined physical spaces. Meanwhile, there is an urgent need for controllable integration of sensing/detecting elements into the reaction channels for rapid diagnosis and feedback of the reaction dynamics to meet the hardware requirements of autonomous artificial intelligence synthesis. However, for most conventional planar microfabrication methods, enabling the extension of the dimensions of microchannel reactors from 2D to 3D with the freedom of tunable footprints and additional functional integration in a facile and cost-effective manner still remains a great challenge. To address this, an alternative 3D microfabrication approach is femtosecond (fs) laser micromachining,[15] which enables high-precision, spatial-selective modification inside transparent materials through non-linear multiphoton absorption for microchannel fabrication and monolithically multifunctional integration. Various types of 3D microchannel structures with on-demand features have been realized by this approach, and show exceptional performance as high-throughput and high-flux microfluidic reactors.[15-17] However, on-chip rapid monitoring of continuous synthesis in 3D multiple microchannels yet has not been realized.[18-20] For instance, 3D facile integration of a probe array (e.g., optical fiber arrays) with high resolutions both in time and space for rapid monitoring and screening of the reaction conditions is highly desirable.[12,14,21]

Here, for the first time to the best of our knowledge, we demonstrate that simultaneous on-chip continuous-flow synthesis and real-time spectroscopic monitoring can be accomplished in a 3D high-performance glass microfluidic chip with an integrated optical fiber probe array. The 3D microfluidic chip consists of three modules: 3D concentration gradient generation, 3D high-efficiency mixing, and on-line spectral monitoring. As a proof-of-concept demonstration, we used the fabricated microfluidic chip system to synthesize zinc oxide (ZnO) nanomaterials, which are an important type of multi-purpose materials due to their high chemical, thermal, and mechanical stability,[22-24] and desirable semiconducting, piezoelectric and optical properties.[25–29] The synthesized ZnO nanomaterial can be tailored in size, shape, and morphology by tuning the flow rates and reactant concentrations based on the spectroscopic signals detected with the fiber probe array.

## Results and discussion

### Device configuration and integration of fiber probe array

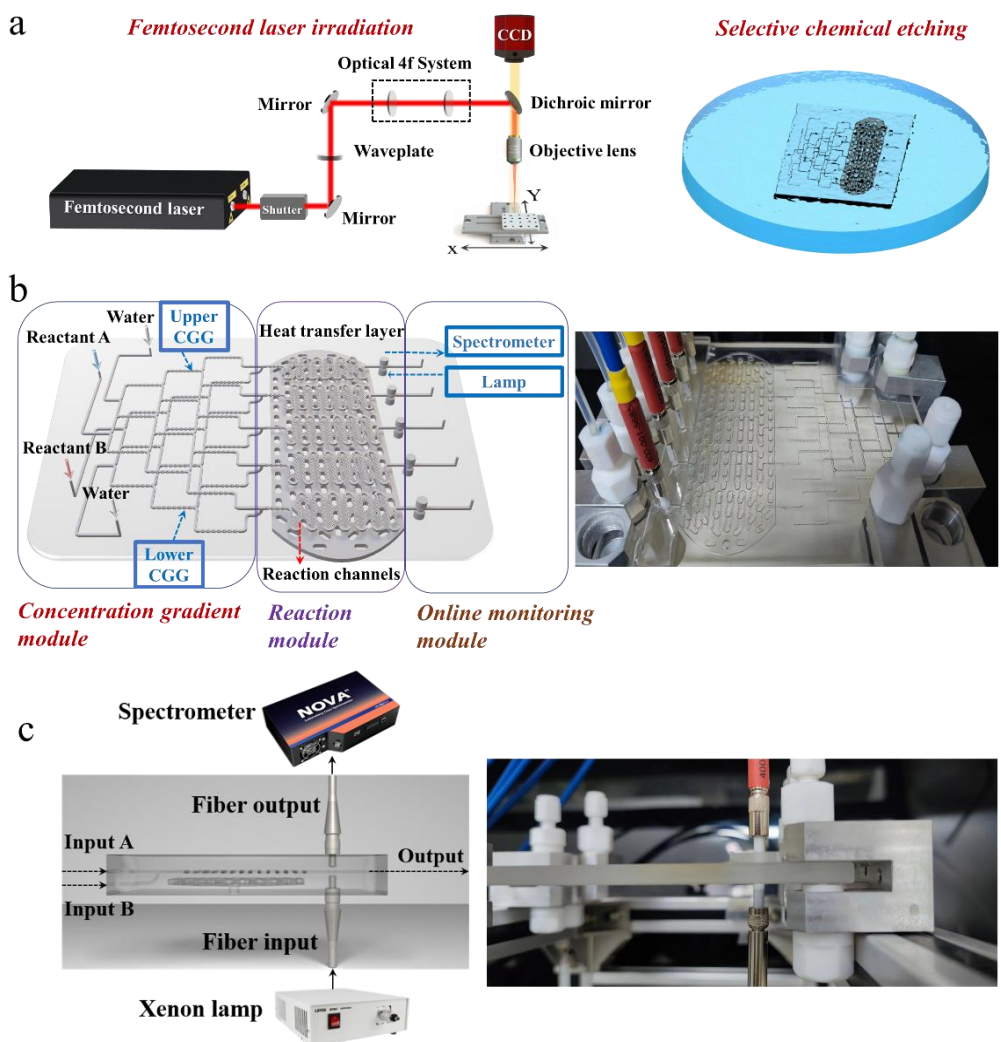

Fig. 1 Schematic illustration of the fabrication process and the microfluidic synthesizer. (a) Fs laser assisted chemical wet etching procedure. (b) Schematic illustration of the microfluidic chip consisting of a concentration gradient module, a reaction module and an on-line monitoring module. (c) Schematic illustration and photograph of the on-line monitoring.

The schematic of the femtosecond laser-assisted etching procedure is illustrated in Fig. 1a (the experimental details are described in the Experimental section). In brief, a piece of the glass substrate was exposed to femtosecond laser irradiation, then sank into a solution of potassium hydroxide (KOH). A microfluidic chip with designed structural features was obtained after removal of the laser-irradiated parts in the glass substrate. The microfluidic chip is of a size of $158 \times 138 \times 9$ mm$^3$ and contains a concentration gradient module, a reaction module and an on-line monitoring module (Fig. 1b).

The concentration gradient module is designed to generate flows with variable concentrations for fast-screening optimized concentrations of reactants.[30, 31] The reaction module consists of five parallel 3D mixing networks sandwiched between the upper and lower layers of heat exchange channels. The mixing networks ensure that the regents are always well-mixed during reaction.[32] The heat exchange channels containing oil inside maintains the temperature of the reactors in the range of –70 ~250 °C during long-term synthesis.

The on-line monitoring module is designed for obtaining spectra with high spatial-temporal resolution. Five microcells are fabricated at the outlet microchannels from the reactors to make sure each reactor can be monitored independently with high spatial resolution. Each microcell contains two round-shaped windows with a diameter of 2.55 mm, matching with the coupled optical fibers. The windows are horizontally arranged with a distance of 1 mm, allowing transmission of light through the microcell. The local flow velocity in the microcells is reduced to minimize influence of the flow velocity on the collected spectra. Considering the characteristic spectra of materials/chemicals are in a wide wavelength range, we select multimode optical fiber (185 nm to 1,100 nm) to couple with the microcells. Due to the same reason, a fiber-coupled xenon lamp (220~2,200 nm LBTEK, LBHPX2000) is utilized as the light source. The wide wavelength range of the optical fiber and xenon lamp indicate that this on-line monitoring module can be applicable to variety of materials/chemicals. The incident light enters the microcell from the bottom window of the microcell. The transmission light from the top window is sent to a fiber-coupled photospectrometer (Idea optics, NOVA 2S) with a wide spectrum range (200~1,100 nm). The photospectrometer is capable of collecting averaged spectra over a short integration time within 10 ms, which strongly guarantee high resolution and precision of the spectra. We should note that shorter integration time leads to higher precision of real-time monitoring, however the final spectral characterization is the average of five individual tests for one sample point to reduce the effect of flow fluctuation and ensure the stability of detection.

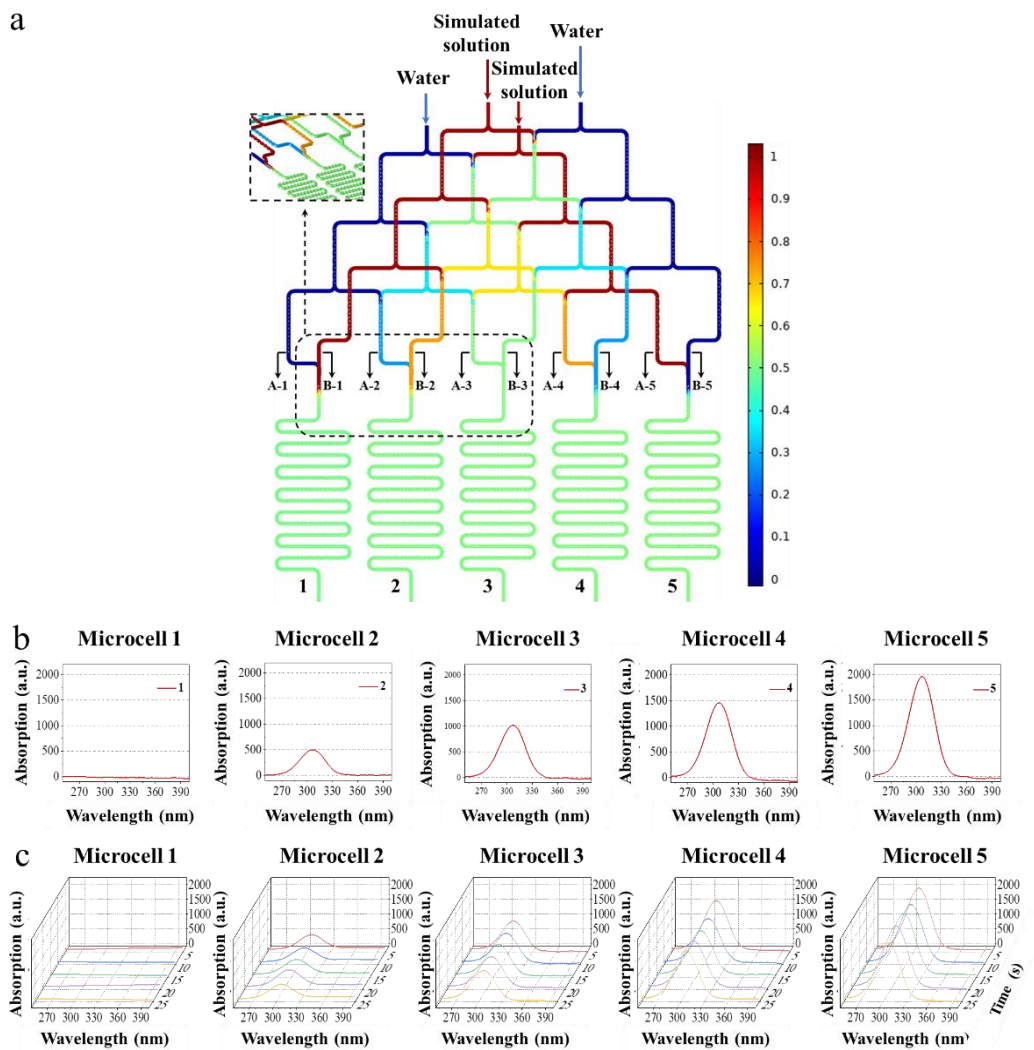

Fig. 2 (a) Numerical simulation results of concentration distribution in the microfluidic chip when the inlet flow rate $Q$ was 12.5 mL min$^{-1}$. (b) Absorption spectra detected at the microcells. (c) Absorption spectra detected at the microcells during a time duration.

**Validation functions of the microfluidic flow synthesizer**

We validated the functions of the integrated microfluidic flow synthesizer before carrying out reactions. Numerical simulation demonstrated that concentration gradient was established among the outlets with an interval of approximately 25% (Fig. 2a). The outlets were subsequently well mixed in the reaction module, resulting in five independent reactors. Please note the details for numerical simulation are described in Supporting Information.

After numerical simulation study, we use on-line monitoring module to collect experimental data to verify the gradient concentration in the reactors and check whether the on-line monitoring module can work properly. Here zinc nitrate solution of a concentration of 300 mM was loaded into the channel via inlets A and B, and deionized (DI) water was loaded via other two inlets with a flow rates of 12.5 mL min$^{-1}$. Fig. 2b illustrates the spectra signal detected at the five microcells. An absorption peak centered at 306 nm was observed in the microcell-2 to microcell-5 except microcell-1 where the fluid should contain DI water only. The absorption corresponds to the presence of $Zn(NO_3)_2$, matching with the absorption spectrum of the sample solutions which were also collected from the outlets measured by a commercial UV-vis spectrophotometer (UV-2600 spectrophotometer, Shimadzu) (Fig. S3).[33] A slight difference between the peak position of the online results (306 nm) and that obtained by the UV-2600 spectrophotometer might be the slight influence of the window or flow fluctuation in the microcell. We plotted a calibration curve based on the intensity of the absorption peak of the spectra of aqueous $ZnNO_3$ solution by using the UV-2600 spectrophotometer. A linear relationship with a coefficient $R^2$ of 0.99 is shown in Fig. S4, suggesting that we can determine the concentration of the $Zn(NO_3)_2$ solution by using UV-vis absorption spectra. In addition, when the flow rate was increased to 25 mL min$^{-1}$, the obtained UV-vis absorption spectra remained similar, indicating that the online monitoring module can work properly under various flow rates in the reactors (Fig. S5). Fig. 2c shows the time-resolved online spectra of the microcells from 5 to 25 seconds. The data were successfully collected. The online spectra collected at each microcell show almost unchanged absorption spectra and peak intensity in the time duration, matching with the steady state of the $Zn(NO_3)_2$ in the reactors. Therefore, the on-line monitoring module with high spatial-temporal resolution was successfully demonstrated.

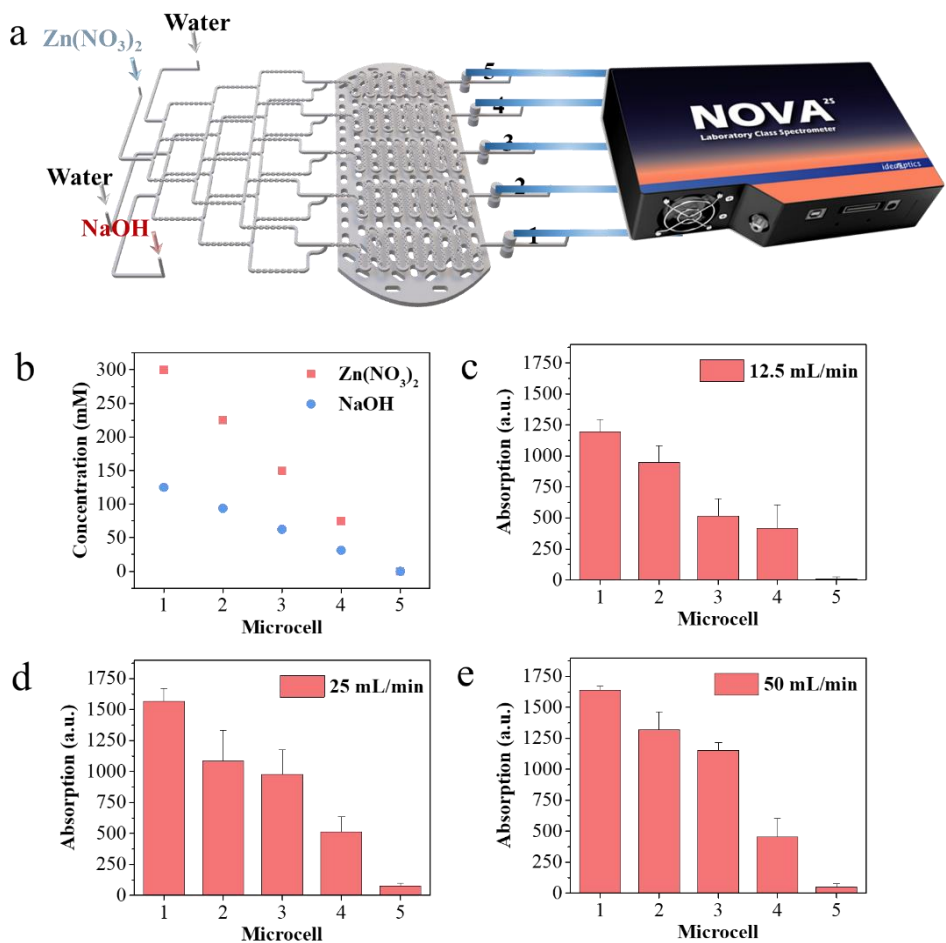

Fig. 3. (a) Schematic illustration for synthesis of ZnO by using the microfluidic synthesizer. (b) Theoretical concentrations of the $Zn(NO_3)_2$ and NaOH solutions which are perfused into the five reactors. (c-e) Absorption intensity at 368 nm extracted from the online spectra. (c) 12.5 mL min$^{-1}$. (d) 25 mL min$^{-1}$. (e) 50 mL min$^{-1}$.

**Flow synthesis of ZnO materials by using the microfluidic flow synthesizer**

The 3D microfluidic synthesizer was used to synthesize ZnO nanostructures. The process is schematically illustrated in Fig. 3a. $Zn(NO_3)_2$ solution (300 mM) and NaOH solution (150 mM) were perfused into microchannels by plunger pump into microchannel at different flow rates. The $Zn(NO_3)_2$ solution in upper channels mixed with the NaOH solution in lower channels through the junction of two layers. The five mixing channels hold solutions with five different concentrations generated from CGGs, resulting in five independent reactions simultaneously on a single chip. Concentration gradient is expected to be generated in the five reactors. The theoretical concentrations of the $Zn(NO_3)_2$ and NaOH solutions which are perfused into the five reactors are shown in Fig. 3b. The reactions were firstly performed under room-temperature. Online monitoring module was employed to record the in-situ spectral signal for preliminary screening. Fig. S6 illustrates the spectra of the products from the five reactors. A singlet strong absorption at 368 nm was found in four reactors except reactor 5 which should not contain $Zn(NO_3)_2$ according to Fig. 3b. The absorption at 368 nm corresponds to 3.38 eV which matches with the band gap of ZnO (3.4 eV in general), corresponding to a wavelength of around 370 nm.[34, 35] To facilitate discussion, we extracted the absorption strength at 368 nm as the signal for ZnO. Fig. 3c presents that ZnO was successfully synthesized in all the reactors except reactor 5 which did not contain $Zn(NO_3)_2$ before reaction. The differences in the absorption strength suggest the difference of yield in all the reactors. In general, the stronger the absorption, the higher the yield of ZnO. We note that the yield linearly correspond to the concentration of $Zn(NO_3)_2$ and NaOH, indicating that the more the reactants, the larger the yield. This trend was kept when the reaction was performed under other flow rates. Fig. 3d-e present that the correlation between the yield of ZnO and the concentration of reactants are almost the same under flow rates of 25 and 50 mL min$^{-1}$. This trend follows the basic rules of chemical reaction that higher concentration of the reactants leads increased collision frequency which results in more products.[34] This observation also suggests that the online monitoring module is suitable for synthesis of ZnO.

This synthesizer offers us a great chance to fast screen a variety of synthetic parameters including the concentration of the reactants and the temperature. The synthetic parameters are changed by tuning flow rates, the concentration gradient of the NaOH solution, $Zn(NO_3)_2$ solution and the temperature of the reactor. Fig. S7 illustrates that 25 samples with varied synthetic parameters were screened. The corresponding online spectra of the 25 samples are listed in Fig. 4. We should note that the whole process for screening all the samples synthesized in a micro reactor array took only one hour excluding the cleaning of the chip, which is much shorter than conventional batch-by-batch synthesis.

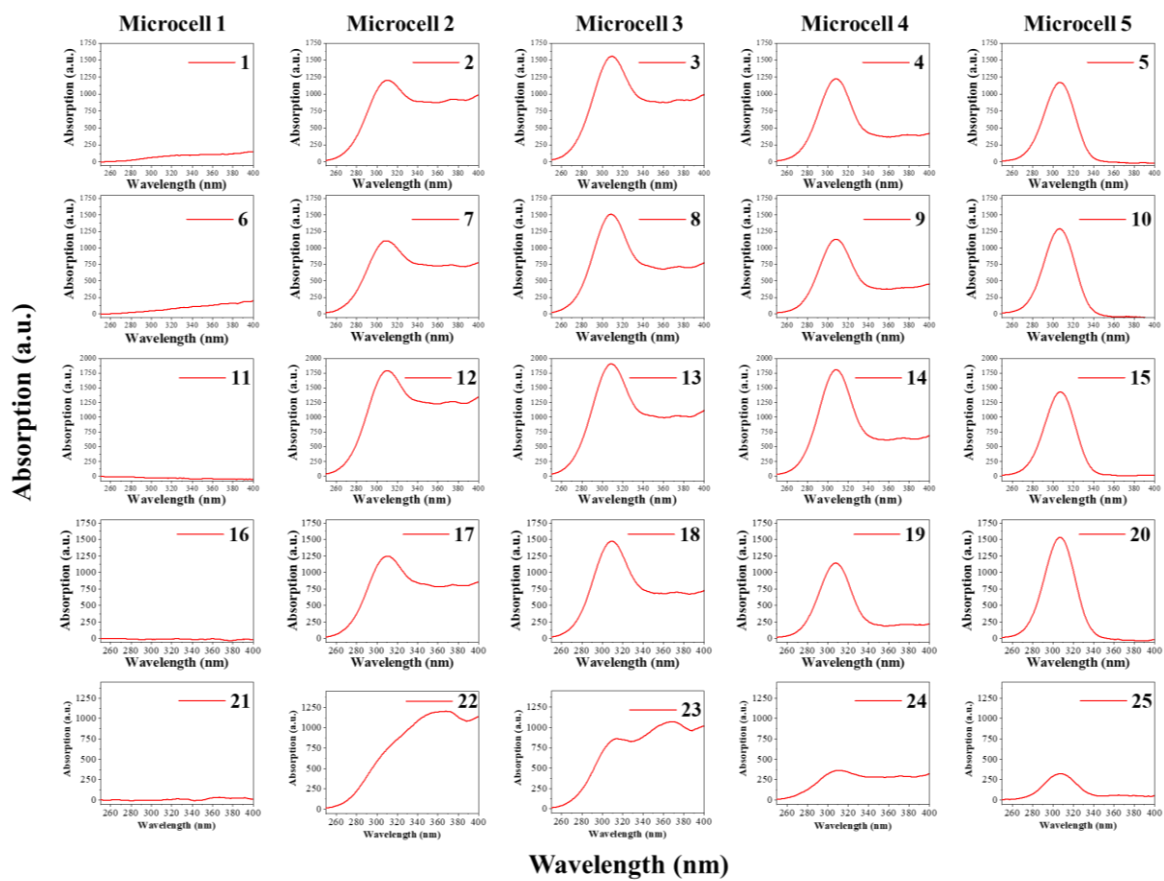

Fig. 4. On-line spectra of the products obtained under varied synthetic conditions as listed in Fig. S7.

Fig. 5a-c illustrate the spectra of three typical samples which were named as sample 12, 22, and 24 according to Fig. 4b. The corresponding synthetic conditions can be found in Fig. 4b. The $I_{306}$ to $I_{368}$ ratios are listed in the blank area in the figures. Fig. 5d-f show SEM images of the selected samples. Sample 12 is composed of microplates with a size range of 150 ~ 400 nm in diameter (Fig. 5d). Sample 22 is composed of flower-like particles with a size of about 10 µm (Fig. 5e). Sample 24 is composed of flower-like particles as well, although the particle size was changed to 1 µm (Fig. 5f). Apparently, the samples are of different sizes and shapes, presenting the power of this microfluidic synthesizer in nanostructure control. To verify the crystal structure of the three samples, HRTEM analyses were taken. Fig. 5g-i present HRTEM image of the particles. Periodic patterns are observed, suggesting that all the particles are crystals. The lattice distance measured in Fig. 5g and Fig. 5i is approximately 2.6 Å, corresponding to lattice distance of {002} facets of plate-like ZnO. On the other hand, the lattice distance measured in Fig. 5h is approximately 2.8 Å, corresponding to lattice distance of {100} facets of flower-like ZnO.[36, 37] Despite

the number of the characterized sample is still small, investigation on the three samples selected by the online spectra can preliminarily suggest that the nanostructure, including the size and shape of the ZnO can probably affect the online spectra. Therefore, on the basis of the online spectra, we may be able to screen the ZnO with the exact nanostructure effectively in the future.

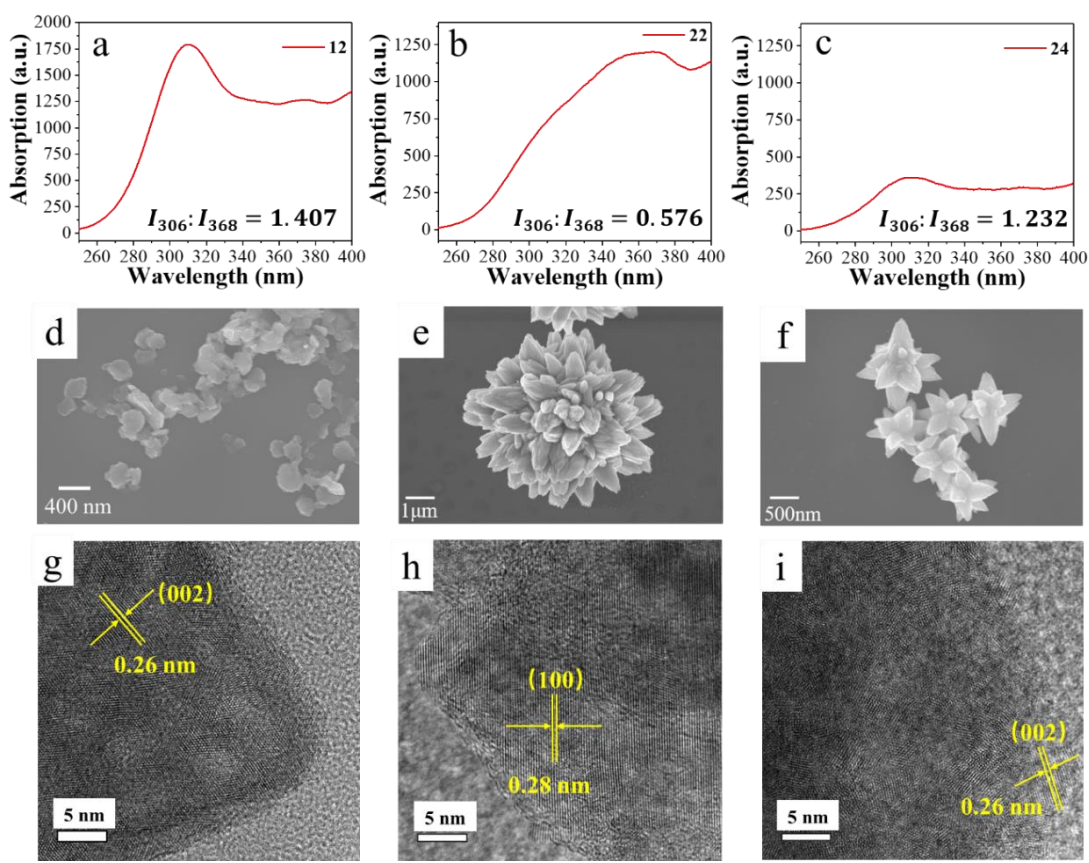

Fig. 5 (a-c) Spectra of three samples which were name as sample 12, 22, and 24, respectively. (d-f) SEM images of sample 12, 22, and 24, respectively. (g-i) HRTEM images of sample 12, 22, and 24, respectively.

**Experimental section**

**Materials**

All the chemicals for on-chip continuous-flow synthesis were used without further purification. Hydrofluoric acid, potassium hydroxide, zinc nitrate, and sodium hydroxide were purchased from Sinopharm Chemical Reagent Co., Ltd. (China).

**Fabrication of the microfluidic chips**

Fabrication of the microfluidic chip was performed by a femtosecond laser micromachining system,[15,38] consisting of two main steps: femtosecond laser irradiation and selective chemical etching by the solution of potassium hydroxide (KOH). To form laser-modified structures, a glass substrate with a size of 158 mm × 138 mm × 9 mm was mounted on a three-dimensional positioning stage (X-axis ABL15020WB, Y-axis ABL15020, Z-axis ANT130-110-L-ZS, Aerotech) and then irradiated by a femtosecond laser source (Pharos, Light Conversion) with programming translation of the stage controlled by a high-performance motion controller (A3200, Aerotech). Prior to the laser irradiation, the spatial configurations of the microfluidic structures were designed and optimized based on numerical modelling results. After laser irradiation, the glass substrate was immersed in an ultrasonic bath of potassium hydroxide with a concentration of 10 M at 95 °C to selectively remove the laser-modified area. Finally, the glass chip containing a concentration gradient module, a reaction module and an on-line monitoring module was cleaned with isopropyl alcohol, deionized (DI) water, and then mildly annealed. Two layers of concentration gradient generator was designed and fabricated integrated with 3D mixers presented in our previous work.[16]

This concentration gradient module consists of two Christmas tree-shaped concentration gradient generators (CGGs), each with 3D mixers designed using Baker's transformation of a unit size of $2 \times 1$ mm$^2$. The two CGGs have a size of 7.2 × 9.8 cm$^2$ each, and located independently in two horizontal layers ahead of the reaction module. The reaction module consists five parallel 3D mixing networks and heat exchange channels. Each of the mixing network is composed of 108 3D mixers which are of unit size of $2 \times 1$ mm$^2$. The heat exchange channels that contains oil inside are of a size of 137.5 × 50 mm$^2$. The oil is able to vary the temperature of the reactors in the range of –70 ~ 250 °C. The online monitoring module consists of five microcells, each microcell containing two round windows with a diameter of 2.55 mm, matching with the coupled multi-mode optical fibers.

**On-chip continuous-flow synthesis**

The on-chip synthesis of ZnO micro- and nano-materials was achieved by introducing $Zn(NO_3)_2$ and NaOH into separate inlets on two layers. The two inlet flows were perfused into microchannels by plunger pumps (custom-made, Sanotac). Concentration gradient was established both for $Zn(NO_3)_2$ and for NaOH. The as-synthesized products were collected at the outlet, and then washed for several times with water, finally got dried at 80 °C.

**Numerical simulation**

The establishment of the concentration gradient was demonstrated by numerical simulation. Here, Reynolds number (Re) was calculated to determine whether the fluids are in laminar flow regimes: Re = $\rho UL/\mu$, where the density ($\rho$ ~1000 kg/ m$^3$) and dynamic viscosity ($\mu$ ~0.001 Pa·s) of water are used for approximations, $L$ is the characteristic length scale of the system, and $U$ is the average flow velocity, which could be obtained by: $U$ = flow rate/wh. In our case with the 3D microchannels, $w$ = 1 mm, $h$ = 1 mm, the flow rate, velocity and Re was shown in Table S1, resulting in laminar flow. We consider them as incompressible with no-slip boundary condition and neglect the gravity force for simplicity. The outlet is set to be fixed pressure with $p$=0 and the diffusion coefficient used is $D = 1\times 10^{-9}$ m$^2$/s.

**Characterization and on-chip monitoring**

The morphology of synthesized products collected from the outlets was characterized by scanning electron microscopy (SEM, Hitachi, S-4800). The crystal structure was examined by high-resolution transmission electron microscopy (HRTEM, JEOL JEM-1400) and X-ray diffraction (XRD Rigaku, Ultima IV, Cu K$\alpha$ radiation). The on-line spectra were obtained by the fiber-coupled photo spectrometer (Idea optics, NOVA 2S) with a fiber-coupled light source xenon lamp (220 nm-2200 nm LBTEK, LBHPX2000) through optical fiber (185 nm to1100 nm) array coupled with the five microcells.

**Conclusions**

In summary, we fabricated a glass microfluidic chip with 3D CGGs and 3D mixers by femtosecond laser-assisted etching. This fabricated microfluidic chip was further integrated with an array of optical fiber probes for on-chip real-time spectroscopic monitoring of the synthesis process with high spatial and temporal resolutions. We demonstrated that ZnO nanostructures could be synthesized with tailored size and shape by tuning the flow rates and reactant concentrations assisted by the feedback spectroscopic signals detected with the fiber probe array. This real-time spectroscopic monitored continuous synthesis can be applied to other molecules and materials to accelerate scientific research and industrial production. The developed femtosecond laser microfabrication technique will pave the way for advanced prototyping of 3D novel glass-based autonomous continuous-flow "factory-on-a-chip" microsystems with intelligent feedback control.

**Funding.** This work was supported by National Natural Science Foundation of China (11734009, 12192251, 11933005，12174107), National Key R&D Program of China (2019YFA0705000), Science and Technology Commission of Shanghai Municipality (21DZ1101500) and Shanghai Municipal Science and Technology Major Project.

**Disclosures.** The authors declare no conflicts of interest.

## Supplementary Information

**Configuration of the concentration gradient (CGGs)**

This concentration gradient module aimed to establish five independent concentration conditions for each reactor. This module consists of two Christmas tree-shaped concentration gradient generators (CGGs), independently in two horizontal layers ahead of the reaction module, each with 3D mixers of a unit size of $2 \times 1$ mm$^2$. The two CGGs have a size of $7.2 \times 9.8$ cm$^2$ each and are located independently in two horizontal layers ahead of the reaction module. The perfused solutions flow through the CGGs and automatically split at bifurcation points to generate outflows with variable concentrations, which can be favourable for screening synthetic parameters of ZnO later. Compared to conventional serpentine microchannels in the CGGs, the 3D mixers designed using Baker's transformation can repeatedly split the fluid into microstreams and reorganize them. The mixers have been shown to possess high mixing efficiency without compromising the flux, while being able to handle flow rates between 5 and 100 mL min$^{-1}$.

**Simulation and analysis of concentration gradient**

Table S1 The parameters of flow dynamics.

| Flow rate (mL/min) | 5 | 12.5 | 25 | 50 |
|---|---|---|---|---|
| $U$ (m/s) | 0.083 | 0.208 | 0.417 | 0.833 |
| Re | 83 | 208 | 417 | 833 |

The governing equations of fluid flow in microfluidic chip channels can be derived from the conservation of mass and conservation of momentum, which are called the continuity equation and Navier-Stokes equation respectively. In the model, it is assumed that the flow state in the microfluidic chip channel is a laminar flow state and steady state, and the internal fluid is an incompressible Newtonian fluid. For incompressible Newtonian fluids, the above equation can be described as:

Equation of continuity:

$$\nabla \times v = 0$$

Momentum equation:

$$\rho \frac{\partial v}{\partial t} + \rho(v \cdot \nabla)v = -\nabla P + \mu \nabla^2 v$$

where $v$ represents the velocity vector, $\rho$ represents the fluid density, $\mu$ represents the dynamic viscosity of the fluid, and P represents the constant temperature pressure. The simulated fluid is set to water, with a viscosity of $1\times10^{-3}$ N*s/m$^2$, and a density of 1000 kg/m$^3$.

Fluid mixing in microfluidic chip channels is mainly controlled by the law of conservation of matter, which is given by the convection and diffusion equation, whose equation can be described as:

Species convection and diffusion equation:

$$\frac{\partial c}{\partial t} + v \cdot \nabla c = D\nabla^2 c$$

c represents the solute concentration; D represents the species diffusion coefficient.

The setting of boundary conditions: at the entrance, the model assumes a fully developed laminar flow, and the velocity of the two entrances is the same as v.

To quantitatively evaluate the mixing efficiency of microfluidic chips I, the formula for calculating the mixing efficiency at any cross-section in the mixing channel is defined as:

$$I = \left(1 - \sqrt{\frac{\sigma_m^2}{\sigma_{max}^2}}\right) \times 100\%$$

$$\sigma_m^2 = \frac{1}{N}\sum_{i=0}^{N}(c_i - \bar{c})^2$$

where N is the total number of a cross-section grid cell, $c_i$ is each grid cell mass concentration, with $\bar{c}$ is $c_i$ for cross-section calculation of average, $\sigma_{max}^2$ is the maximum variance of mixture concentration along each cross-section of the channel (that is, the variance of concentration when there is no mixing at the entrance). $\sigma_{max}^2$ is calculated by the following formula:

$$\sigma_{max}^2 = \bar{c}(1 - \bar{c})$$

The value of mixing efficiency I varies from 0 (no mixing) to 100% (full mixing).

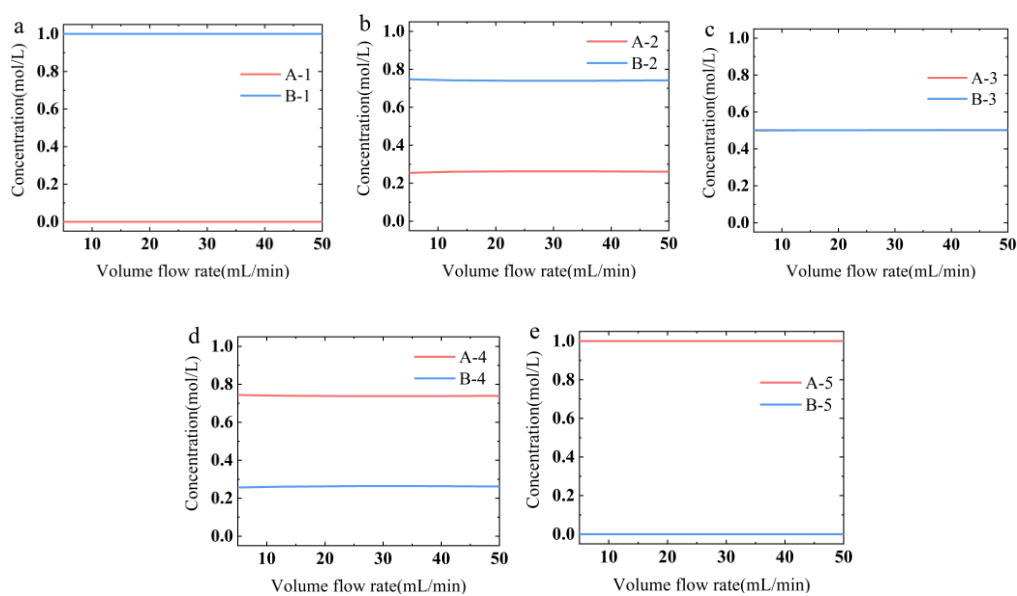

**Fig.S1** Concentrations of the fluids at outlets before entering the reaction module at different flow rates.

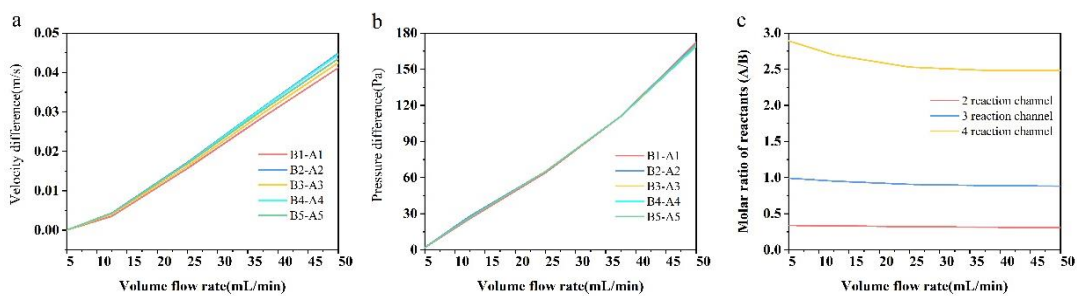

**Fig. S2** Numerical simulation results at the outlet ahead of the reaction module at different flow rates. (a) Velocity difference. (b) Pressure difference. (c) Molar ratio.

In this work, two kinds of fluid were simulated, one with solute, and the other without solute, to facilitate normalization. The fluids at the outlets which are named A1 to A5 in the lower layer and B1 to B5 in the upper layer show different colors, indicating a concentration gradient interval of approximately 25%. The outlets were subsequently mixed in the reaction module. The uniform color shown in the reaction channel has an obvious mixing effect, realizing complete mixing quickly after entering.

In microfluidic flow synthesis, the flow rate is an important synthetic parameter that can determine the molar ratio of matter and may influence the reaction kinetics. Therefore, we use numerical simulation to consider the influence of the flow rate on the concentration gradient, the velocity difference, and the pressure difference at the two entrances of each reactor. Fig. S1 shows that the concentration gradient was almost steady in a range of flow rates from 5 to 50 mL min$^{-1}$. Fig. S2a presents that the velocity difference at the two entrances of each reactor gradually increased with the flow rate. This is probably caused by the increased pressure difference as shown in Fig. S2b. The molar ratio of substances in channels 2, 3, and 4 were calculated (Fig. S2c). With the increase in the flow rate, the molar ratio in all three reaction channels showed a downward trend and eventually stabilized. The absence of any crossing of molar ratios indicates that the reaction conditions remained in a step distribution. Therefore, in the following case experiments, the reactions carried out are all stepped conditions within the range of the flow rate tested.

**Validation functions of the microfluidic flow synthesizer**

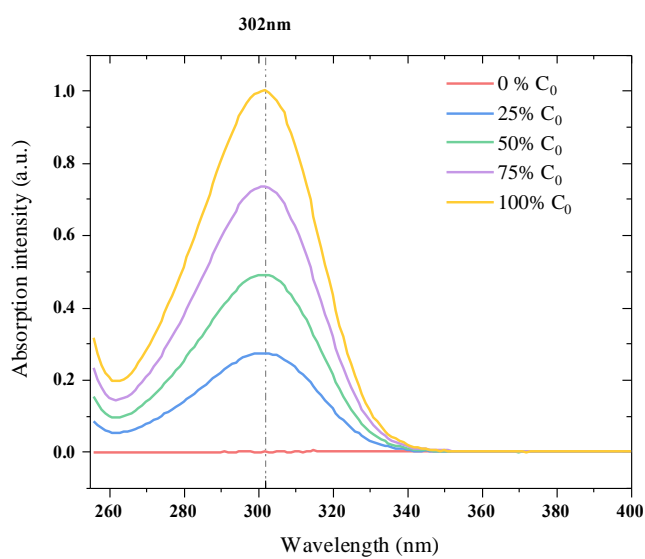

Fig. S3 Ultraviolet-visible absorption spectrum of zinc nitrate solution measured by a commercial UV-2600 photo spectrometer. $C_0$ represents the initial concentration of $Zn(NO_3)_2$.

After the numerical simulation study, we use an online monitoring module to collect experimental data to verify the gradient concentration in the reactors and check whether the online monitoring module can work properly. Here zinc nitrate solution of a concentration of 300 mM was loaded into the channel via inlets A and B, and deionized (DI) water was loaded via the other two inlets with a flow rate of 12.5 mL min$^{-1}$.

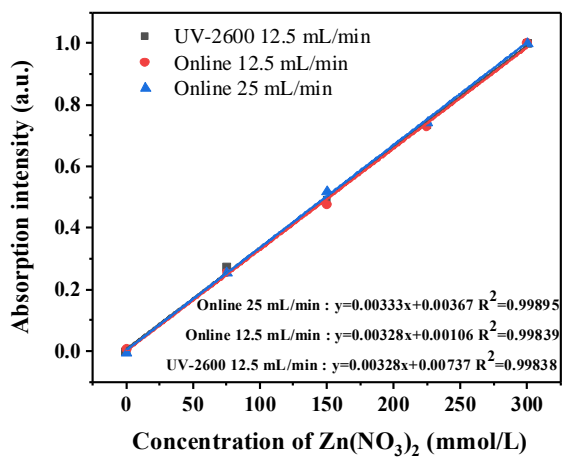

Fig. S4. Correlation between the concentration of $Zn(NO_3)_2$ and the peak intensity as measured by an online monitoring module and a commercial UV-2600 photo spectrometer.

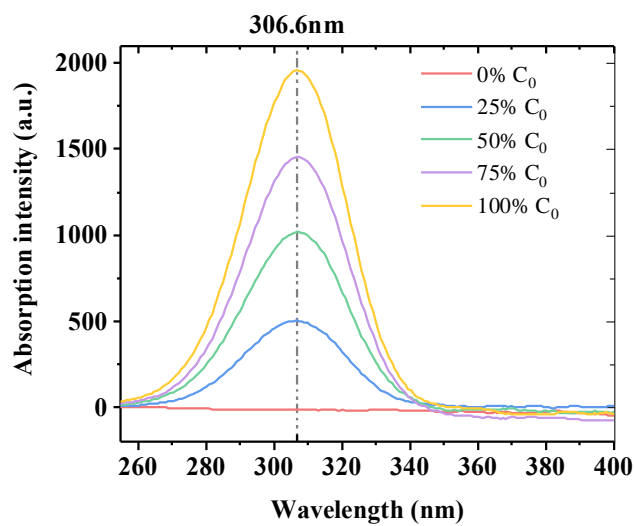

Fig. S5. Online spectra taken at the microcells when the inlet flow rate $Q$ was 25 mL min$^{-1}$. $C_0$ represents the initial concentration of $Zn(NO_3)_2$.

We extracted the peak intensities of the online spectra and analysed the relationship between the peak intensities and the calculated concentration of the $Zn(NO_3)_2$ in the microcells by assuming that the concentration gradient with a gradient interval of approximately 25% established in the microcells. A linear relationship that was very similar to the calibration curve was observed in Fig. S4, confirming the generation of the concentration gradient and the accuracy and the high spatial resolution of the online spectra. In addition, when the flow rate was increased to 25 mL min$^{-1}$, the online spectra presented a very close linear relationship between the concentration of the $Zn(NO_3)_2$ and the peak intensity (with a coefficient R2 of 0.99 again) (Fig. S4 and S5). Therefore, the online monitoring module can work properly under varied flow rates in the reactors.

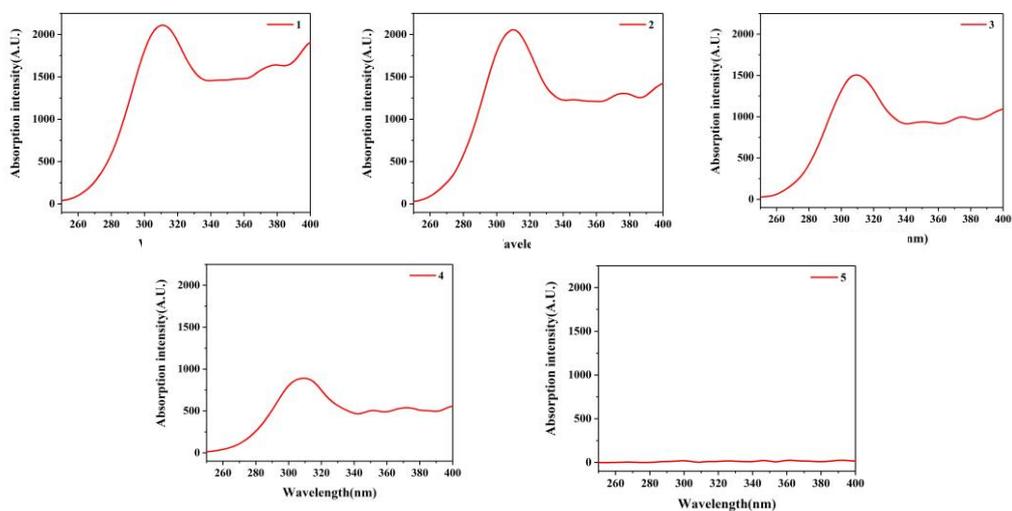

Fig. S6. Online spectra of the products obtained under room temperature with a fluidic rate of 12.5 mL/min.

| 1 | 2 | 3 | 4 | 5 |
|---|---|---|---|---|
| NaOH :125 mmol/L<br>ZnNO$_3$: 0 mmol/L<br>V=25 mL/min<br>T=25 °C | NaOH :93.75 mmol/L<br>ZnNO$_3$: 75 mmol/L<br>V=25 mL/min<br>T=25 °C | NaOH :62.5 mmol/L<br>ZnNO$_3$: 150 mmol/L<br>V=25 mL/min<br>T=25 °C | NaOH :31.25 mmol/L<br>ZnNO$_3$: 225 mmol/L<br>V=25 mL/min<br>T=25 °C | NaOH : 0 mmol/L<br>ZnNO$_3$: 300 mmol/L<br>V=25 mL/min<br>T=25 °C |
| 6 | 7 | 8 | 9 | 10 |
| NaOH :125 mmol/L<br>ZnNO$_3$: 0 mmol/L<br>V=12.5 mL/min<br>T=25 °C | NaOH :93.75 mmol/L<br>ZnNO$_3$: 75 mmol/L<br>V=12.5 mL/min<br>T=25 °C | NaOH :62.5 mmol/L<br>ZnNO$_3$: 150 mmol/L<br>V=12.5 mL/min<br>T=25 °C | NaOH :31.25 mmol/L<br>ZnNO$_3$: 225 mmol/L<br>V=12.5 mL/min<br>T=25 °C | NaOH : 0 mmol/L<br>ZnNO$_3$: 300 mmol/L<br>V=12.5 mL/min<br>T=25 °C |
| 11 | 12 | 13 | 14 | 15 |
| NaOH :125 mmol/L<br>ZnNO$_3$: 0 mmol/L<br>V=25 mL/min<br>T=90 °C | NaOH :93.75 mmol/L<br>ZnNO$_3$: 75 mmol/L<br>V=25 mL/min<br>T=90 °C | NaOH :62.5 mmol/L<br>ZnNO$_3$: 150 mmol/L<br>V=25 mL/min<br>T=90 °C | NaOH :31.25 mmol/L<br>ZnNO$_3$: 225 mmol/L<br>V=25 mL/min<br>T=90 °C | NaOH : 0 mmol/L<br>ZnNO$_3$: 300 mmol/L<br>V=25 mL/min<br>T=90 °C |
| 16 | 17 | 18 | 19 | 20 |
| NaOH :125 mmol/L<br>ZnNO$_3$: 0 mmol/L<br>V=12.5 mL/min<br>T=90 °C | NaOH :93.75 mmol/L<br>ZnNO$_3$: 75 mmol/L<br>V=12.5 mL/min<br>T=90 °C | NaOH :62.5 mmol/L<br>ZnNO$_3$: 150 mmol/L<br>V=12.5 mL/min<br>T=90 °C | NaOH :31.25 mmol/L<br>ZnNO$_3$: 225 mmol/L<br>V=12.5 mL/min<br>T=90 °C | NaOH : 0 mmol/L<br>ZnNO$_3$: 300 mmol/L<br>V=12.5 mL/min<br>T=90 °C |
| 21 | 22 | 23 | 24 | 25 |
| NaOH :125 mmol/L<br>ZnNO$_3$: 0 mmol/L<br>V=5 mL/min<br>T=90 °C | NaOH :93.75 mmol/L<br>ZnNO$_3$: 15 mmol/L<br>V=5 mL/min<br>T=90 °C | NaOH :62.5 mmol/L<br>ZnNO$_3$: 30 mmol/L<br>V=5 mL/min<br>T=90 °C | NaOH :31.25 mmol/L<br>ZnNO$_3$: 45 mmol/L<br>V=5 mL/min<br>T=90 °C | NaOH : 0 mmol/L<br>ZnNO$_3$: 60 mmol/L<br>V=5 mL/min<br>T=90 °C |

Fig. S7. List of the synthetic parameters of the 25 screened samples.